\documentclass[prd,superscriptaddress,amsfonts,amssymb,twocolumn,amsmath,showpacs]{revtex4-2}
\usepackage{bm}
\usepackage{amsfonts}
\usepackage{tensor}

\usepackage{latexsym}
\usepackage[utf8]{inputenc}
\usepackage{graphicx}
\usepackage{amsmath}
\usepackage{palatino}
\usepackage{mathpazo}
\usepackage[british]{babel}
\usepackage{hhline}
\usepackage{multirow}
\usepackage{textcomp}
\usepackage{float}
\usepackage{booktabs}
\usepackage{dcolumn}
\usepackage{lipsum} 
\usepackage{mdframed}
\usepackage[]{mdframed}
\usepackage{hhline}
\usepackage{multirow}
\usepackage{ragged2e}
\usepackage{hyperref}
\hypersetup{colorlinks,citecolor=blue}
\hypersetup{colorlinks=true,linkcolor=red,filecolor=magenta,    urlcolor=cyan}
\usepackage{amsmath}
\usepackage{xcolor}
\usepackage{orcidlink}
\usepackage{epsfig}
\usepackage{caption}
\usepackage{subcaption}
\usepackage{commath}

\def\jnl@style{\it}
\def\aaref@jnl#1{{\jnl@style#1}}

\def\aaref@jnl#1{{\jnl@style#1}}

\def\aj{\aaref@jnl{AJ}}                   
\def\apj{\aaref@jnl{ApJ}}                 
\def\apjl{\aaref@jnl{ApJ}}                
\def\apjs{\aaref@jnl{ApJS}}               
\def\apss{\aaref@jnl{Ap\&SS}}             
\def\aap{\aaref@jnl{A\&A}}                
\def\aapr{\aaref@jnl{A\&A~Rev.}}          
\def\aaps{\aaref@jnl{A\&AS}}              
\def\mnras{\aaref@jnl{Mon.~Not.~Roy.~Astron.~Soc.}}             
\def\prd{\aaref@jnl{Phys.~Rev.~D}}        
\def\prc{\aaref@jnl{Phys.~Rev.~C}}  
\def\prl{\aaref@jnl{Phys.~Rev.~Lett.}}    
\def\qjras{\aaref@jnl{QJRAS}}             
\def\skytel{\aaref@jnl{S\&T}}             
\def\ssr{\aaref@jnl{Space~Sci.~Rev.}}     
\def\zap{\aaref@jnl{ZAp}}                 
\def\nat{\aaref@jnl{Nature}}              
\def\aplett{\aaref@jnl{Astrophys.~Lett.}} 
\def\apspr{\aaref@jnl{Astrophys.~Space~Phys.~Res.}} 
\def\physrep{\aaref@jnl{Phys.~Rep.}}      
\def\physscr{\aaref@jnl{Phys.~Scr}}       
\def\commat{\aaref@jnl{Comm.~Math.~Phys.}}              
\def\science{\aaref@jnl{Science}}               
\def\cqg{\aaref@jnl{Classical Quant.~Grav.}}            
\def\jpcs{\aaref@jnl{JPCS}}                                     
\def\ijmpd{\aaref@jnl{Int.~J.~Mod.~Phys.~D}}                    
\def\grg{\aaref@jnl{Gen.~Relat.~Gravit.}}               
\def\rpp{\aaref@jnl{Rep.~Prog.~Phys.}}          
\def\npa{\aaref@jnl{Nucl.~Phys.~A}}        
\def\lrr{\aaref@jnl{Living Rev.~Rel.}}                   
\def\jcap{\aaref@jnl{J.~Cosmology Astropart.~Phys.}}    
\def\rmp{\aaref@jnl{Rev.~Mod.~Phys.}}   
\def\epjc{\aaref@jnl{Eur.~Phys.~J.~C}} 
\def\plb{\aaref@jnl{~Phy.~Lett.~B}} 
\def\mpla{\aaref@jnl{Mod.~Phy.~Lett.~A}} 
\def\arxiv{\aaref@jnl{arxiv.org}}

\allowdisplaybreaks[1]

\begin{document}

\color{black}       

\title{Cosmological Dynamics of Hyperbolic Evolution Models in $f(Q,L_m)$ Gravity}

\author{V. A. Kshirsagar\orcidlink{0009-0003-1256-2246}}
\email{kvitthal99@gmail.com}
\affiliation{Shri Jagdishprasad Jhabarmal Tibrewala University, Vidhya Nagari, Churu-Jhunjhunu Road, Chudela, Jhunjhunu, Rajasthan 333010}
\author{S. A. Kadam\orcidlink{0000-0002-2799-7870}}
\email{siddheshwar.kadam@dypiu.ac.in;
\\k.siddheshwar47@gmail.com}
\affiliation{Centre for Interdisciplinary Studies and Research, D Y Patil International University, Akurdi, Pune-411044, Maharashtra, India}
\author{Vishwajeet S. Goswami}
\email{vishwajeetgoswami.math@gmail.com}
\affiliation{Shri Jagdishprasad Jhabarmal Tibrewala University, Vidhya Nagari, Churu-Jhunjhunu Road, Chudela, Jhunjhunu, Rajasthan 333010}
\begin{abstract}
\textbf{Abstract}: This paper highlights cosmologically viable sine and cosine hyperbolic evolution functions in the framework of $f(Q,\mathcal{L}_m)$ gravity. The models have been tested to check the behavior of the equation of state (EoS) parameter under the variation of parametric values. The EoS parameter experiences a quintessence phase, and is approaching to $-1$ at late time. The models are showing inclined behaviour with the $\Lambda$CDM model at the late time. The viability of both the models is retested using the widely accepted energy conditions in both cases. The violation of the strong energy condition admits the accelerating behaviour of the models.  The same has been explained through the analysis of the profile of deceleration parameter, which concretely supports the evidence that the models explain early deceleration to late time acceleration phenomena.
\end{abstract}
\maketitle

\section{Introduction}\label{sec:I}
Over the last couple of decades, advances in observational astronomy have led to the discovery that the Universe is not only expanding but doing so at an accelerating rate. A observation firstly identified in the late 1990s by supernova measurements \cite{Riess_1998, Perlmutter_1999, Riess_2004}, and confirmed by baryon acoustic oscillations \cite{Eisenstein_2005} and cosmic microwave background anisotropies \cite{Planck_2020}. The study of accelerating phenomena have become one of the main focuses of modern cosmology. This phenomena have been generally modeled through a dark energy (DE) component comprising nearly about $70\%$ of the total energy budget of the universe. This can be described through a cosmological constant within the standard $\Lambda$CDM model. However, these frameworks still face major theoretical challenges-mostly the fine-tuning and coincidence problems that arise in linking with the cosmological constant \cite{Peebles:2002gy, Weinberg_1989}. To address this issue, the modified gravity theories are playing the promising role. One way to study geometric deep learning  for cosmic acceleration is to modifying general relativity (GR). The initial extensions of GR is $f(R)$ gravity \cite{Sotiriou_2010}, and it's further modifications can be analysed into scalar–tensor theories \cite{Fujii:2003pa} and Gauss–Bonnet modifications \cite{Nojiri_2005}. These modifications changes the gravitational action in order to include additional curvature terms or new dynamic scalar degrees of freedom. Although these curvature-based modifications flesh out the landscape of viable cosmological models, but these modifications are of higher-order field equations \cite{Stelle:1976gc, Sotiriou_2010} or require invisible mechanisms to avoid instabilities and to be consistent with solar system tests \cite{Clifton_2012, Nojiri_2011}.

A promising and relatively uncharted approach has arisen from the geometric formulation of gravity that incorporates nonmetricity scalar $Q$. In symmetric teleparallel gravity, the Levi-Civita connection is supplanted by a flat connection that is torsion-free, with the sole geometric characteristic being nonmetricity \cite{Nester:1998mp,BeltranJimenez_2017}. The work of J. Beltr{\'a}n Jim{\'e}nez et al derives the full background and perturbation equations and discusses the emergence of additional scalar degrees of freedom. This work serves as a fundamental guide for cosmological applications of modified gravity based on non-metricity \cite{BeltranJimenez_2019}. Successive developments of modified gravities, from curvature based modification to teleparallel and symmetric teleparallel schemes, have progressively enlarged the geometric interpretation of gravity. It involves the interactions beyond curvature to include torsion and non-metricity. From the starting the importance is gained by one of the initial symmetric teleparallel gravity theory $f(Q)$ gravity in light of its novelty in approach to the gravitational interface via the non-metricity scalar $Q$ \cite{Heisenberg_2023}. The recent observation studies have been performed in $f(Q)$ gravity formalism \cite{Kadam:2026fq}. This formalism have successfully developed further incorporating the boundary term $C$ \cite{De:2024fQC,Samaddar:2025mxw},  and the generalization $f(Q,\mathcal{L}_m)$ gravity, with the inclusion of the matter Lagrangian $\mathcal{L}_m$, introduces a non-minimal coupling between geometry and matter, which finds significant implications in the dynamics of cosmology.

Observational data representative of late-time cosmic acceleration and, among other tensions, the Hubble parameter discrepancy underlines the practical relevance of pursuit of the $f(Q, \mathcal{L}_m)$ gravity theories \cite{Myrzakulov:2024kuu}.  The particular problem addressed by this review is the formulation and analysis of energy conditions within  $f(Q,\mathcal{L}_m)$ gravity under several scale factor progresses which administrate the expansion history of the universe. The observational constraints on  $f(Q,\mathcal{L}_m)$ gravity theories have been analysed in \cite{Myrzakulov:2024esv}. The studies focus on the non-conservation of the matter energy-momentum tensor due to geometry-matter coupling, important for arguments about the physical sustainability and thermodynamic consistency of these models \cite{Hazarika:2024alm, Alfedeel:2024ktc}. Different methodologies for demonstrating the matter Lagrangian and its coupling with non-metricity produce different expectations for cosmic acceleration and stability, underlining the need for systematic evaluation \cite{Myrzakulov:2024twj, ZeeshanGul:2025fky}. Since the combined analysis of energy conditions mixed with scale factor variation parameters stands as a gap, it results in a particular constraint on the predictive power of $f(Q,\mathcal{L}_m)$ gravity in cosmology \cite{Bekkhozhayev:2024qvq}. The modified Friedmann equations obtained from $f(Q,\mathcal{L}_m)$ gravity summarize this interaction, while energy conditions (null, weak, strong, and dominant) help as criteria for physical viability and cosmic evolution \cite{Hazarika:2024alm}.

This paper is organized into six sections, Section \ref{sec:I} provides an introduction and a general overview of the research problem, Section \ref{sec:II} presents the theoretical background of symmetric teleparallel gravity and the formulation of $f(Q,\mathcal{L}_m)$ theory. Section \ref{sec:III} derives the field equations for a spatially flat FLRW spacetime and formulates the effective thermodynamic quantities. Section \ref{sec:IV} analyzes specific models and parameter constraints, Section \ref{sec:V} discusses the energy conditions and their application to $f(Q,\mathcal{L}_m)$ gravity. Finally, Section \ref{conclusion} summarizes the main findings, highlights the physical implications, and outlines possible directions for future research.

\section{Overview of $f(Q,\mathcal{L}_m)$ Gravitational Theory}
\label{sec:II}
The affine connection \(Y^{\alpha}{}_{\mu\nu}\) in Weyl–Cartan geometry is exceptionally distributed into three essential parts: the symmetric Levi-Civita connection \(\Gamma^{\alpha}{}_{\mu\nu}\), the contortion tensor \(K^{\alpha}{}_{\mu\nu}\) that captures the antisymmetric feature, and the disformation tensor \(L^{\alpha}{}_{\mu\nu}\), which designates the non-metricity. So, it is usually written as follows \cite{Xu:2019sbp} :
\begin{equation}
   Y^{\alpha}{}_{\mu\nu} =\Gamma^{\alpha}{}_{\mu\nu} + K^{\alpha}{}_{\mu\nu} + L^{\alpha}{}_{\mu\nu}\,.
\end{equation}
The torsion-free Levi-Civita connection \(\Gamma^{\alpha}{}_{\mu\nu}\) is identical with the second-order Christoffel symbols derived from the metric tensor \( g_{\mu\nu} \). It conserves the inner products of tangent vectors during parallel transport, hence it is metric-compatible. This connection is clearly stated by the standard formula in terms of the metric and its partial derivatives, it is expressed according to 
\begin{equation}
\Gamma^{\alpha}{}_{\mu\nu}
= \frac{1}{2}\, g^{\alpha\lambda}
\left( 
\partial_{\mu} g_{\lambda\nu}
+ \partial_{\nu} g_{\lambda\mu}
- \partial_{\lambda} g_{\mu\nu}
\right)\,,
\end{equation}
\( K^{\alpha}{}_{\mu\nu} \) is the contortion tensor. it represents a dissimilarity between a general affine connection with torsion and the torsion-free Levi-Civita connection of the metric. It is constructed from the torsion tensor \( T^{\alpha}{}_{\mu\nu} \), taking the antisymmetric deviation in the connection coefficients. This tensor encrypts how torsion changes parallel transport and geodesics in spacetime geometries away from standard general relativity.
\begin{equation}
    K^{\alpha}{}_{\mu\nu} = \frac{1}{2} \left(T^{\alpha}{}_{\mu\nu} + T_{\mu}{}^{\alpha}{}_{\nu} + T_{\nu}{}^{\alpha}{}_{\mu}\right)\,.
\end{equation}
 The disformation tensor \(L^{\alpha}{}_{\mu\nu}\), by encoding the effects of non-metricity inside the connection, allows vector length to not necessarily be conserved under parallel transport. As a matter of fact, this involvement will be relevant for geometric frameworks where the metric is not preserved under displacement, hence spreading spacetime structure outside its standard Riemannian picture. Explicitly,
\begin{equation}
    L^{\alpha}{}_{\mu\nu} = \frac{1}{2} \left(Q^{\alpha}{}_{\mu\nu} - Q_{\mu}{}^{\alpha}{}_{\nu} - Q_{\nu}{}^{\alpha}{}_{\mu} \right)\,.
\end{equation}
Concerning the Weyl–Cartan connection \(Y^{\alpha}{}_{\mu\nu}\), The non-metricity tensor quantifies the departure from metric compatibility, describing how the metric concludes to remain unaffected under covariant differentiation. It is expressed as
\begin{equation}
    Q_{\alpha\mu\nu} = \nabla _ {\alpha} g_{\mu\nu}=\partial_{\alpha} g_{\mu\nu} - Y^{\beta}{}_{\alpha\mu}\, g_{\beta\nu} - Y^{\beta}{}_{\alpha\nu}\, g_{\mu\beta}\,.
\end{equation}
In order to construct a boundary term in the action of metric-affine gravity theories, we consider the conjugate of the non-metricity tensor, called superpotential \( P^{\alpha}{}_{\mu\nu} \), defined as \cite{DAgostino:2018ngy}:
\begin{equation}
    P^{\alpha}{} _ {\mu\nu} =  \frac{-1}{2} {L}^{\alpha}{}_{\mu\nu} + \frac {1}{4} \left({Q}^{\alpha} - \tilde {Q} ^ {\alpha} \right) g_{\mu\nu} - \frac {1}{4} {\delta}^{\alpha} ({}_{\mu}Q_{\nu})\,.
\end{equation}
Where, \(Q^{\alpha} = Q^{\alpha}{}_{\mu}{}^{\mu}\) and \(\tilde{Q}^{\alpha} = Q_{\mu}{}^{\alpha\mu}\) are two independent non-metricity vectors. By contraction of the superpotential tensor with the non-metricity tensor, the related non-metricity scalar can be achieved: 
\begin{equation}
    Q = - Q_{\lambda\mu\nu}\ P^{\lambda\mu\nu}\,.
\end{equation}
In this work, we examine an extension of symmetric teleparallel gravity, The action is expressed on the framework $f(Q,\mathcal{L}_m)$ modified gravity as \cite{Hazarika:2024alm}:
\begin{equation}\label{frl1}
    S=\int \left[\mathcal{L}_m + f(Q,\mathcal{L}_m) \right] \sqrt{-g} d^4x.
\end{equation}
where, $\sqrt{-g}$ signifies the determinant of the metric tensor. The function $f(Q,\mathcal{L}_m)$ arbitrarily depends on both $Q$ non-metricity scalar and $\mathcal{L}_m$ matter Lagrangian. This structure enlarges the theoretical background, covering beyond the standard originations of symmetric teleparallel gravity and general relativity (GR). To derive equivalent equations to the gravitational field equations, the action given by \eqref{frl1} needs to be varied with respect to the metric tensor \(g_{\mu\nu}\). This leads to the field equations given below:
\begin{equation}\label{Fe.Eq.}
\begin{aligned}
\frac{2}{\sqrt{-g}}\nabla_{\alpha}\!\left(
f_{Q}\sqrt{-g}\, P^{\alpha}{}_{\mu\nu}
\right)
&+ f_{Q}\left(
P_{\mu\alpha\beta} Q_{\nu}{}^{\alpha\beta}
- 2 Q^{\alpha\beta}{}_{\mu} P_{\alpha\beta\nu}
\right) \\
&+ \frac{1}{2} f\, g_{\mu\nu}
= \frac{1}{2} f_{\mathcal{L}_m}
\left(
g_{\mu\nu} \mathcal{L}_m - T_{\mu\nu}
\right).
\end{aligned}
\end{equation}
In this framework, the derivatives of the function with respect to its arguments are represented by $f_{Q} = \frac{\partial f(Q,\mathcal{L}_m)}{\partial Q}$ and $f_{\mathcal{L}_m} = \frac{\partial f(Q,\mathcal{L}_m)}{\partial\mathcal{L}_m}$. A appropriate interpretation arises when the function factors as $f(Q,\mathcal{L}_m) = f(Q)+2\mathcal{L}_m$; in this development, the field equations are reduced to the conventional ones given by an $f(Q)$ gravity, as shown in \cite{BeltranJimenez_2017}. The energy–momentum tensor $T_{\mu\nu}$ for the matter sector can be obtained by varying the action of the matter sector with respect to the metric tensor. It can be known as-
\begin{equation}
    T_{\mu\nu} = -\frac{2}{\sqrt{-g}}\frac{\delta\left(\sqrt{-g} \mathcal{L}_m\right)}{\delta g^{\mu\nu}}= g_{\mu\nu} \mathcal{L}_m - 2 \frac{\partial \mathcal{L}_m}{\partial g^{\mu\nu}}.
\end{equation}
A self-determining variation of the action with respect to the affine connection outcomes in further dynamical equation: 
\begin{equation}
    \nabla_{\mu} \nabla_{\nu}\left(4 \sqrt{-g}\, f_{Q}\, P^{\mu\nu}{}_{\alpha}+ H_{\alpha}{}^{\mu\nu}\right)= 0 \,.
\end{equation}
Here, \(H_{\alpha}{}^{\mu\nu}\) denotes the hypermomentum density, which describes the clear dependence of the matter Lagrangian on the affine connection. It is presented by the explanation:
\begin{equation}
   H_{\alpha}{}^{\mu\nu}= \sqrt{-g} {f_\mathcal{L}}_m\frac{\delta \mathcal{L}_m}{\delta Y^{\alpha}{}_{\mu\nu}} \,.
\end{equation}
In this expression, the existence of the hypermomentum term decides this framework from purely metric theories and hence, essentially establishes its metric–affine characteristic. A characteristic of $f(Q,\mathcal{L}_m)$ gravity is that the energy–momentum tensor is normally not conserved. Actually, taking the covariant derivative of the field equations obtained previously \eqref{Fe.Eq.}, one achieves the following relation:
\begin{equation}
\begin{aligned}
D_{\mu} T^{\mu}{}_{\nu}
&= \frac{1}{f_{\mathcal{L}_m}}
\Bigg[
\frac{2}{\sqrt{-g}} \nabla_{\alpha} \nabla_{\mu}
H_{\nu}{}^{\alpha\mu}
+ \nabla_{\mu} A^{\mu}{}_{\nu} \\
&\hspace{2.2em}
- \nabla_{\mu}\!\left(
\frac{1}{\sqrt{-g}} \nabla_{\alpha}
H_{\nu}{}^{\alpha\mu}
\right)
\Bigg] = B_{\nu} \neq 0 \,.
\end{aligned}
\end{equation}
This relation just makes apparent that the energy–momentum tensor $T^{\mu}{}_{\nu}$ does not satisfy the normal conservation law condition, declaring that energy and momentum are not conserved in the usual sense. Vector $B_{\nu}$ computes this non-conservation and stems from the coupling between matter fields and the nonmetric part of the geometry. Its form depends on the specific $f(Q,\mathcal{L}_m)$ chosen, on the comprehensive structure of the matter Lagrangian $\mathcal{L}_m$, and on the dynamics of the additional fields present in the theory.

\section{Friedmann equations for Modified  $f(Q,\mathcal{L}_m)$ Gravity} \label{sec:III}
To study the energy conditions in the context of $f(Q,\mathcal{L}_m)$ gravity, we will describe the universe by means of a spatially flat Friedmann–Lemaître–Robertson–Walker (FLRW) spacetime. In this cosmological description we assume two essential hypotheses concerning the large-scale structure of the universe \cite{Ryden_2016}, the homogeneity and isotropy FLRW background, the geometry of spacetime is described by the following line element:
\begin{equation}
    ds^{2} = -dt^{2} + a^{2}(t)\left(dx^{2} + dy^{2} + dz^{2}\right)\,.
\end{equation}
Here, $a(t)$ is the cosmological scale factor whose time evolution describes the expansion of the Universe. In the context of this cosmological situation, the non-metricity scalar reduces to $Q = 6H^{2}$, where the Hubble parameter $H=\frac{\Dot{a}}{a}$ describes the rate of the Universe's expansion. Additionally, it is assumed that the universe is modeled by a perfect fluid. This gives a matter Lagrangian density of $\mathcal{L}_m=-\rho$ \cite{Samaddar:2025qxr}, which is equivalent to $\mathcal{L}_m=\rho$ \cite{Harko:2015pma}. The energy-momentum tensor, when described in the comoving coordinate system, has a diagonal form of $T^{\mu}{}_{\nu}=(-\rho,p,p,p)$. The matter part is defined by assuming that the energy-momentum tensor is that of a perfect fluid, which is characterized by the following tensor:
\begin{equation}
    T_{\mu\nu} = (\rho + p) u_{\mu}u_{\nu} + p g_{\mu\nu}\,.
\end{equation}
Here, $\rho$ denotes the energy density, $p$ is the isotropic pressure, and $u_\mu$ represents the four-velocity of the fluid. Using the FLRW metric in the equations of the fields, expressed in Eq. $\eqref{Fe.Eq.}$, two Friedmann equations can be obtained \cite{Hazarika:2024alm}- 
\begin{equation}\label{Eq.16}
    3H^{2} = \frac{1}{4 f_{Q}}\left[f - f_{\mathcal{L}_m}\left(\rho + \mathcal{L}_m \right)\right]\,,
\end{equation}
\begin{equation}\label{Eq.17}
   \dot{H} + 3H^{2} + \frac{\dot{f}_{Q}}{f_{Q}} H = \frac{1}{4 f_{Q}} \left[f+f_{\mathcal{L}_m} \left(p-\mathcal{L}_m \right)\right]\,.
\end{equation}   
In particular, when considering a function of the form $f(Q,\mathcal{L}_m) = f(Q) + 2\mathcal{L}$, a set of generalized Friedmann equations equivalent to those derived in $f(Q)$ gravity can be obtained, which will be used to recover the symmetric teleparallel equivalent of General Relativity. The cosmological and solar system tests have strongly supported the General Relativity theory. Therefore, since any theoretical deviation needs to be small, this implies a linear form of $f(Q)$. Therefore, based on this idea and with the intent of studying the evolution of the universe with viscosity, we will adopt a function in the following form \cite{Hazarika:2024alm}:
\begin{equation}
    f(Q,\mathcal{L}_m)=\frac{-Q}{2} + \alpha Q^{\mu} \mathcal{L}_m +\beta\,.
\end{equation}
Within this framework, the value of $\mu$ is considered a free parameter along with the values of $\alpha$ and $\beta$. Here, the value of $\mu$ had been chosen to be unity to have a linear coupling between the fields $Q$ and $\mathcal{L}_m$; in this given model, the value of $\mu$ will vary. The assumption made in this model allows the dimensionality of the parameters to be expanded in the process of investigating the effect of the exponent of the nonminimal coupling on the evolution of the universe. Consequently, it follows directly that $f_Q=\alpha  \mu Q^{\mu -1}\mathcal{L}_m-\frac{1}{2}$ and $f_{\mathcal{L}_m}=\alpha  Q^{\mu }$. For this specific form of this function, using $\mathcal{L}_m=\rho$ \cite{Harko:2015pma}, From the relation of the Hubble parameter with the scale factor, $H=\frac{\dot{a}}{a}$, the non-metricity scalar simplifies to $Q=6H^2$. Then, from equations \eqref{Eq.16} and \eqref{Eq.17} one may obtain an explicit expression for the effective equation-of-state (EoS) parameter that enables the analysis of the dark energy phase of cosmic evolution:
\begin{equation} \label{Eq.19}
    \omega=\frac{-2 \dot{H} \left[\beta  \mu +3 H^2 (1+\mu) \right]-3 H^2 (\beta +3 H^2)}{3 H^2 (\beta +3 H^2)}\,.
\end{equation}
To close and solve the system of equations, it is necessary to impose a suitable relation involving the matter sector or the metric functions. In the FLRW framework, spacetime is isotropic and characterized by identical metric potentials; therefore, in the following section, we adopt a specific, well-motivated form of the scale factor to proceed with the analysis.

\section{Cosmological Models} \label{sec:IV}
To examine the role of the cosmological background dynamics, we investigate the field equations \eqref{Eq.16} and \eqref{Eq.17} along with the equation-of-state parameter described by equation \eqref{Eq.19}. As these expressions can be cast solely in terms of Hubble parameter $H$, it is helpful to choose a functional form of $H$.

\subsection{Model-I $a(t) = \gamma \cosh{\gamma t}$}\label{m1} 
\begin{figure}[H]
    \centering
\includegraphics[width=0.45\textwidth]{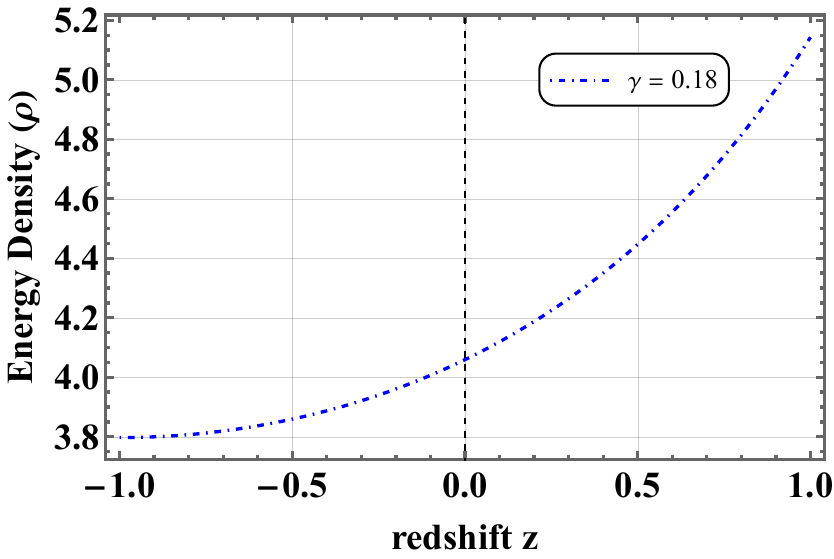}
    \caption{Evolution of energy density as functions of redshift $z$ for $\alpha=0.18, \mu=1.3, \beta=0.5, \gamma=0.18$.}
    \label{fig:energy-density:m1}
\end{figure}

Let us assume the first cosmic scale factor $a(t) = \gamma \cosh{\gamma t}$ \cite{Kadam:2022lxt, Singh:2022gln}, where $\gamma$ is an arbitrary constant. The corresponding hyperbolic evolution of the Hubble parameter $H(t)$ described by $H(t) = \gamma \tanh{(\gamma t)}$. Then, by referring to the conventional relation between the redshift and the scale factor $a=\frac{1}{1+z}$, the Hubble parameter $H(z)$ can be transformed into $H(z) = \gamma \sqrt{1-(1+z)^{2} \gamma ^2}$. Thus deceleration parameter can be transformed to be described solely by the redshift $z$ as follows:
\begin{equation}
    q(z)= -1 + \frac{(1+z) \gamma}{\left[1-(1+z)^2 \gamma ^2 \right]^{3/2}}\,.
\end{equation}
Since, these expressions are highly complicated, it becomes difficult to study the characteristics of the parameters analytically. To understand the variation of physical parameters, we use graphical analysis. The dynamics of the parameters in our model is largely dependent on the choice of parameters related to the scale factor and on the choice of the $f(Q,\mathcal{L}_m)$ function. For an accelerating model, it has to be ensured that the energy density is always positive, and then the pressure turns negative with the negative equation of state at our time and at larger times. Based on these physical considerations, we choose values of parameters related to the scale factor. Since our scale factor varies with one free parameter, we study our model with some illustrative choices of parameters.The energy density is always greater than zero during the evolution of the cosmos for various values of the parameter $\gamma$ as shown in Fig. \ref{fig:energy-density:m1}. 
In these scenarios, the energy density has a monotonically decreasing behavior from the early universe to the late universe.

\begin{figure}[ht]
    \centering
    \includegraphics[width=0.45\textwidth]{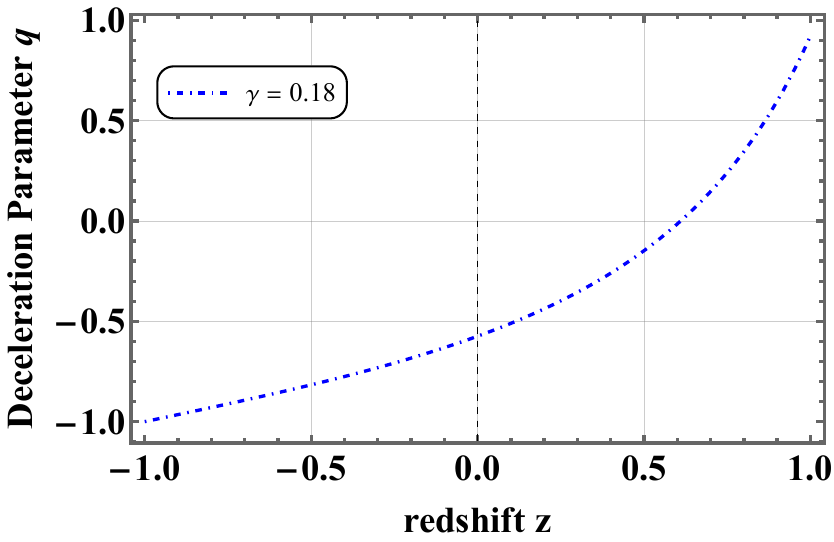}
    \caption{Evolution of deceleration parameter as functions of redshift $z$ for $\gamma=0.18$.}
    \label{fig:deceleration:m1}
\end{figure}

The deceleration parameter characterizes how the universe is expanding. A positive deceleration parameter $q$ indicates a slowing expansion, akin to the matter-dominated epoch, whereas a negative $q$ points to an accelerated expansion characteristic of the DE-dominated phase. As illustrated in Fig. \ref{fig:deceleration:m1}, the current deceleration parameter is approximately $-0.52 \pm 0.06$ \cite{Capozziello:2019DE}. This reinforces the notion of the ongoing accelerated expansion of the universe, with the negative value of $q_0$ signifying the current predominance of dark energy. Additionally, the transition happens at a redshift of $z_{tr} \approx 0.52$, which is in line with the commonly accepted transition redshift derived from Planck-CMB 2018 observations within the context of the $\Lambda$CDM model~\cite{Planck:2018vyg}.

\begin{figure}[H]
    \centering
    \begin{subfigure}[t]{0.45\textwidth}
        \centering
        \includegraphics[width=\textwidth]{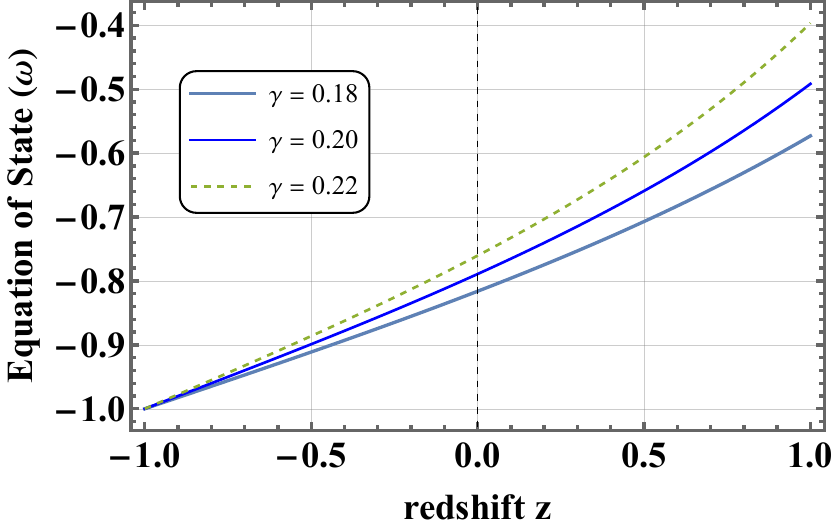}
        \caption{EoS parameter with changing $\gamma$}
        \label{fig:Eos:gamma}
    \end{subfigure}
    \begin{subfigure}[t]{0.45\textwidth}
        \centering
        \includegraphics[width=\textwidth]{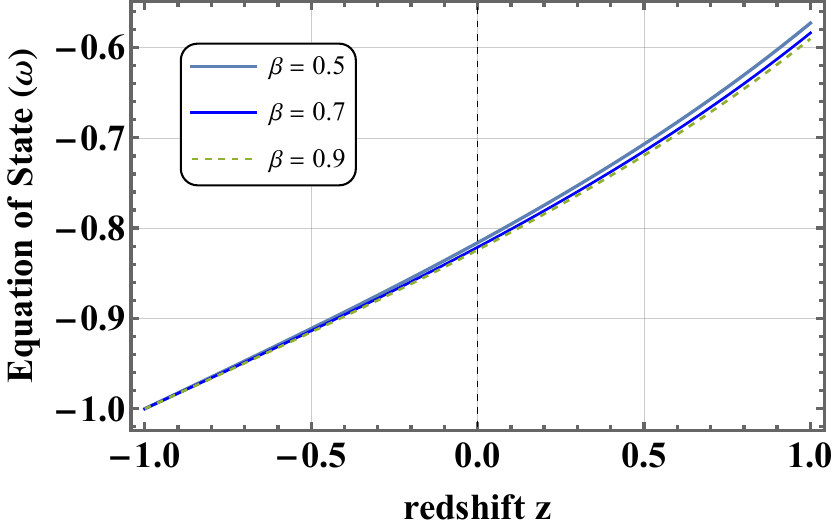}
        \caption{EoS parameter with changing $\beta$}
        \label{fig:Eos:beta}
    \end{subfigure}
    \begin{subfigure}[t]{0.45\textwidth}
        \centering
        \includegraphics[width=\textwidth]{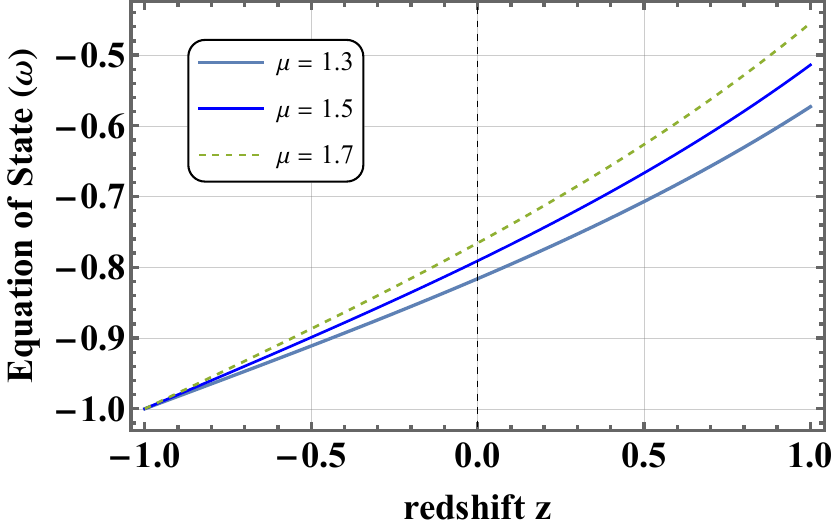}
        \caption{EoS parameter with changing $\mu$}
        \label{fig:Eos:mu}
    \end{subfigure}
    \caption{Evolution of the EoS parameter with respect to redshift $z$ for the parametric values mentioned in the caption of Fig. \ref{fig:energy-density:m1}.}
    \label{EOSm1}
\end{figure}

The behaviour of energy density $\rho$ in Fig. \ref{fig:energy-density:m1} is showing decreasing behaviour from early to late time. The equation of state (EoS) parameter $\omega$ defines the characteristics of the universe's expansion. There are three potential scenarios for the universe's acceleration: the cosmological constant where $\omega=-1$, the phantom regime where $\omega < -1$, and the quintessence regime where $-1<\omega<-1/3$. Fig. \ref{EOSm1} shows that $\omega$ stays above $-1$ across the redshift range and tends to approach $\omega \approx -1$ as we move further into the future. This pattern suggests that the dark energy (DE) in our model operates within a quintessence regime with $\omega_{0} = -0.80 \pm 0.4$, aligning with current observational studies~\cite{Capozziello:2014}. The EoS parameter is examined by varying different model parameters. It has been observed that the behavior of the EoS parameter is not significantly affected by changes in the $\beta$ parameter, as it remains within the current quiescence region. Additionally, the parameter takes on higher values when the parameters $\gamma$ and $\mu$ are increased.

\subsection{Model-II $a(t)=\sinh(\eta t)^{\frac{1}{n}}$}\label{m2}
Let us consider the second cosmic scale factor $a(t)= \sinh(\eta t)^{\frac{1}{n}}$ \cite{Mishra:2020saj, Chawla:2012lpe}, where $\eta$ and $n$ are arbitrary constants. The corresponding hyperbolic evolution of the Hubble parameter $H(t)$ described by $H(t) = \frac{\eta  \coth (\eta  t)}{n}$. Then, by referring to the conventional relation between the redshift and the scale factor $a=\frac{1}{1+z}$, the Hubble parameter $H(z)$ can be transformed into $H(z) = \frac{\eta}{n} \sqrt{1+(1+z)^{2 n}}$. In this case, the deceleration parameter will take the form as follow,
\begin{equation}
\begin{aligned}
q(z)
&= -1
-\frac{n^2 \left(\frac{1}{z+1}\right)^{2 n}}%
{\eta^2 \left[\left(\frac{1}{z+1}\right)^{2 n}+1\right]}
\Bigg(
\eta \left(\frac{1}{z+1}\right)^{1-n}
\Bigg. \\
&\hspace{2.8em}
\Bigg.
\sqrt{\left(\frac{1}{z+1}\right)^{2 n}+1}- \frac{\eta \left(\frac{1}{z+1}\right)^{n+1}}%
{\sqrt{\left(\frac{1}{z+1}\right)^{2 n}+1}}
\Bigg) \,.
\end{aligned}
\end{equation}

\begin{figure}[H]
    \centering
    \includegraphics[width=0.45\textwidth]{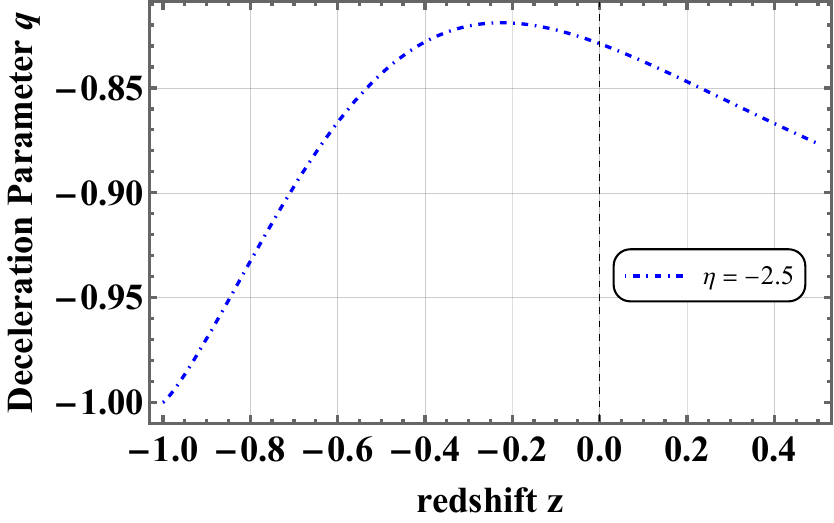}
    \caption{Evolution of deceleration parameter as functions of redshift $z$ for $\eta=-2.5$.}
    \label{fig:deceleration:m2}
\end{figure}

\begin{figure}[H]
    \centering
    \includegraphics[width=0.45\textwidth]{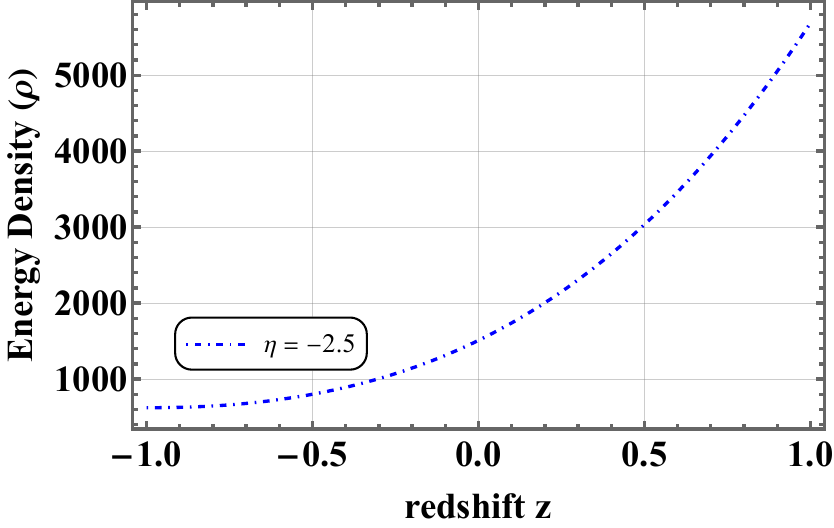}
    \caption{Evolution of energy density as functions of redshift $z$ for $\alpha=0.18, \mu=-0.3, \beta=0.5$, $n=1.1$, $\eta=-2.5$.}
    \label{fig:energy-density:m2}
\end{figure}

The deceleration parameter presented in Fig. \ref{fig:deceleration:m2} lies in the negative region for the current and the late time, which explains the current acceleration. The values of $q_{0}=-0.80 \pm 0.2$ this is consistent with the observational studies presented in \cite{Capozziello:2019DE}. With these settings, the expression of energy density pressure and the EoS parameter presented in \eqref{Eq.16}, \eqref{Eq.17}, \eqref{Eq.19} can be transformed to be described solely by the redshift $z$ as follows:

\begin{figure}[H]
    \centering
    \begin{subfigure}[t]{0.45\textwidth}
        \centering
        \includegraphics[width=\textwidth]{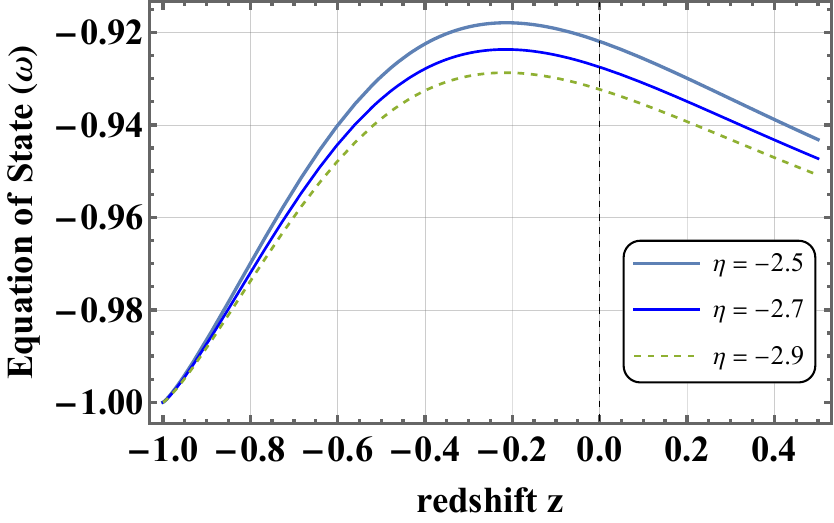}
        \caption{EoS parameter with changing $\eta$}
        \label{fig:Eos:eta}
    \end{subfigure}
    \begin{subfigure}[t]{0.45\textwidth}
        \centering
        \includegraphics[width=\textwidth]{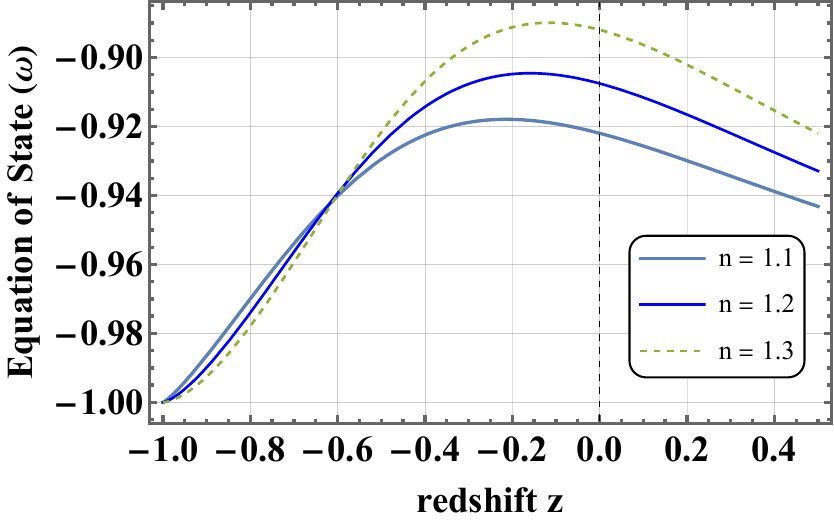}
        \caption{EoS parameter with changing $n$}
        \label{fig:Eos:beta}
    \end{subfigure}
    \begin{subfigure}[t]{0.45\textwidth}
        \centering
        \includegraphics[width=\textwidth]{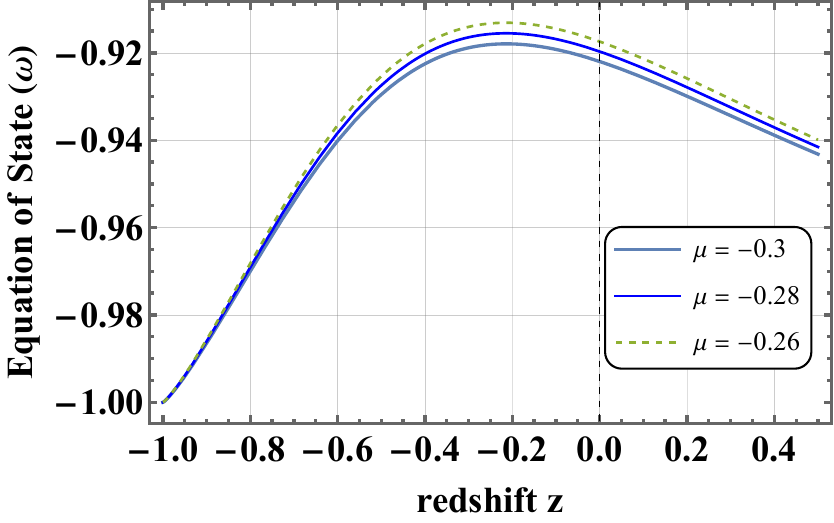}
        \caption{EoS parameter with changing $\mu$}
        \label{fig:Eos:mu}
    \end{subfigure}
    \caption{Evolution of the EoS parameter with respect to redshift $z$ for the parametric values mentioned in the caption of Fig. \ref{fig:energy-density:m2}.}
    \label{EoS:m2}
\end{figure}

The energy density shown in Fig. \ref{fig:energy-density:m1} for Model-\ref{m2} exhibits higher values than Model-\ref{m1}. Additionally, the energy density demonstrates a decreasing trend. The EoS parameter has been analyzed for various model parameters and is presented in Fig. \ref{EoS:m2}. The EoS parameter lies within the quintessence region, approaching -1 in the late time, which aligns with the behavior of the $\Lambda$CDM model. The plots indicate that, at the present time, the EoS parameter takes on higher values for larger model parameters, specifically $\eta$, $n$, and $\mu$. The present value $\omega_{0}=-0.9 \pm 0.2$ which aligns with the observational studies in \cite{Adame:2025DESI, Brout:2022PantheonPlus}. 

\section{Energy Conditions}\label{sec:V}
\begin{figure}[H]
    \centering
     \includegraphics[width=0.45\textwidth]{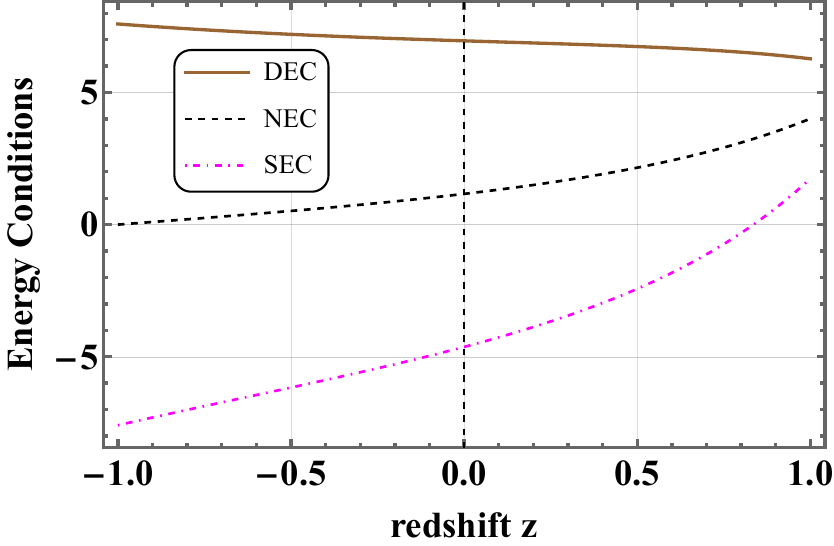}
    \includegraphics[width=0.46\textwidth]{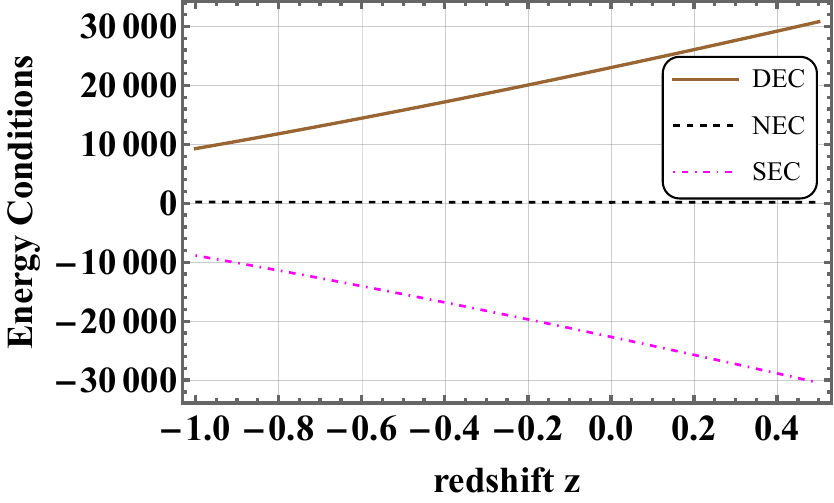}
    \caption{Evolution of energy conditions for Model I: Dominant (DEC), Strong (SEC), and Null (NEC) as functions of redshift $z$ for $\alpha=0.18, \mu=1.3, \beta=0.5, \gamma=0.18$ for Model I and  $\alpha=0.18, \mu=-0.3, \beta=0.5, \eta=-0.3, n=1.1$ for Model II respectively.}
    \label{fig:energy-conditions}
\end{figure}

Energy conditions impose restrictions on the energy-momentum tensor that are consistent across different coordinate systems. The main conditions are \cite{Raychaudhuri_1955, Carroll_2019, Curiel_2017,Bhagat:2023ych}:  
\begin{itemize}
    \item \textbf{Weak Energy Condition (WEC):}  
    $T_{ij}t^{i}t^{j}\geq 0$ for every timelike vector $t^{i}$.  
    For a perfect fluid:  
    \begin{equation}
        T_{ij}u^{i}u^{j}=\rho, \quad 
        T_{ij}\xi^{i}\xi^{j}=(\rho+p)(u_{i}\xi^{i})^{2}.
    \end{equation}
    This indicates that $\rho \geq 0$ and $\rho+p \geq 0$.  
    \item \textbf{Null Energy Condition (NEC):}  
    $T_{ij}\xi^{i}\xi^{j}\geq 0$ for any null vector $\xi^{i}$.  
    This condition is equivalent to $\rho+p \geq 0$.  
    \item \textbf{Strong Energy Condition (SEC):}  
    $T_{ij}t^{i}t^{j}-\tfrac{1}{2}T^{k}_{~k}t^{l}t_{l}\geq 0$.  
    This is equivalent to $\rho+p \geq 0$ and $\rho+3p \geq 0$.  
    It suggests that gravity acts attractively.  
    \item \textbf{Dominant Energy Condition (DEC):}  
    $T_{ij}t^{i}t^{j}\geq 0$ and $T^{ij}t_{i}$ is non-spacelike.  
    For a perfect fluid: $\rho \geq |p|$.  
\end{itemize}
The energy conditions fundamentally act as the boundary conditions that shape the development of the universe and are illustrated in Fig. \ref{fig:energy-conditions}. Furthermore, owing to the inherent structure of spacetime, these energy conditions structurally describes the gravitational attraction. The NEC remains positive throughout the cosmic journey, indicating that the energy density is non-negative. The DEC is consistently maintained, guaranteeing that the energy density exceeds the pressure and that energy transmission adheres to causal limits. The SEC is violated during the early and later stages, which corresponds with the observed accelerated expansion and suggests a deviation from general matter-dominated frameworks. Evaluating these conditions allows us to ascertain the properties of matter and energy within the Universe, which is essential for understanding its accelerated expansion and the role of dark energy.

\section{Summary and Conclusion}\label{conclusion}
In this study, we explored two well-established hyperbolic evolution functions within the modified symmetric teleparallel \( f(Q, \mathcal{L}_m) \) gravity formalism. We investigated the specific form of the function given by 
$f(Q, \mathcal{L}_m) = \frac{-Q}{2} + \alpha Q^{\mu} \mathcal{L}_m + \beta$ \cite{Hazarika:2024alm}. We calculated the field equations for this setup, which are presented in Eqs. \eqref{eq:prhoomegam1} and \eqref{eq:prhoomegam2}. Our analysis included a study of the behavior of energy density, the equation of state (EoS) parameter, and the deceleration parameter. We tested energy conditions to assess the viability of the models. The energy density exhibited a decreasing trend in both models, as shown in Figs. \ref{fig:energy-density:m1} and \ref{fig:energy-density:m2}. We also examined the EoS parameter for varying values of the parameters, with results displayed in Figs. \ref{EOSm1} and \ref{EoS:m2}. In both cases, the EoS lies within the quintessence region, approaching \(-1\) at late times, which is consistent with the \(\Lambda\)CDM model. The current values for the EoS parameters are \(\omega_{0} = -0.80 \pm 0.4\) and \(\omega_{0} = -0.9 \pm 0.2\), which align with observational studies referenced in \cite{Adame:2025DESI, Brout:2022PantheonPlus}.

We further investigated the behavior of the deceleration parameter, which is crucial for understanding the contributions from the evolution function. The deceleration parameter for the first model is presented in Fig. \ref{fig:deceleration:m1}, indicating a transition from early deceleration to late-time cosmic acceleration, with a value of \(q_{0} = -0.52 \pm 0.06\). In contrast, the deceleration parameter for Model \ref{m2} remains negative, explaining the current acceleration phase of the universe's evolution, with \(q_{0} = -0.80 \pm 0.2\), which is compatible with recent observational studies cited in \cite{Capozziello:2019DE}. To reaffirm the cosmological viability of these models, we demonstrated the energy conditions, with plots presented in Fig. \ref{fig:energy-conditions}. In both models, the NEC is satisfied, indicating that the energy density remains positive. The DEC is also upheld in both cases, suggesting that the energy density exceeds the pressure throughout the evolution. However, the violation of the SEC supports the accelerating epochs of the universe's evolution. The best fit cosmological parameters results along with energy conditions analysis have been summerised in the Table \ref{Table_1}.

\begin{table}[ht]
\centering
\label{tab:models}
\begin{tabular}{|l|c|c|c|}
\hline
\textbf{Quantity} & \textbf{Model I} & \textbf{Model II} & \textbf{Consistent with} \\
\hline
$w_0$ & $-0.8 \pm 0.4$  & $-0.9 \pm 0.2$  & Refs.\cite{Adame:2025DESI, Brout:2022PantheonPlus} \\
\hline
$q_0$ & $-0.52 \pm 0.06$ & $-0.8 \pm 0.2$ & Ref. \cite{Capozziello:2019DE} \\
\hline
NEC & Satisfied & Satisfied &
\multirow{3}{*}{\begin{tabular}{c}
Late-time\\
acceleration
\end{tabular}} \\
\cline{1-3}
DEC & Satisfied & Satisfied &  \\
\cline{1-3}
SEC & Violated  & Violated  &  \\
\hline
\end{tabular}
\caption{Best-fit cosmological parameters and energy condition analysis for the considered models.}\label{Table_1}
\end{table}

\section{Appendix}
Equations for Model-\eqref{m1}, \eqref{m2} are presented respectively as follows in Eqs \eqref{eq:prhoomegam1}, \eqref{eq:prhoomegam2} respectively,
\begin{widetext}
\begin{equation}\label{eq:prhoomegam1}
\begin{aligned}
p(z) ={}&
\frac{2^{-\mu} 3^{-1-\mu} \gamma^{-2(1+\mu)}
\left[1-(1+z)^{2}\gamma^{2}\right]^{-1-\mu}}
{\alpha(1+2\mu)}
\Bigg[
-3\gamma^{2}
\left[1-(1+z)^{2}\gamma^{2}\right]
\left(
\beta + 3\gamma^{2}
\left[1-(1+z)^{2}\gamma^{2}\right]
\right)
\\[-0.3em]
&\qquad
+ \frac{
2(1+z)\gamma^{3}
\left(
\beta\mu
+ 3\gamma^{2}(1+\mu)
\left[1-(1+z)^{2}\gamma^{2}\right]
\right)}
{\sqrt{1-(1+z)^{2}\gamma^{2}}}
\Bigg],
\\[0.6em]
\rho(z) ={}&
\frac{
6^{-\mu} \gamma^{-2\mu}
\left[1-(1+z)^{2}\gamma^{2}\right]^{-\mu}
}
{\alpha(1+2\mu)}
\left(
\beta
+ 3\gamma^{2}
\left[1-(1+z)^{2}\gamma^{2}\right]
\right),
\\[0.6em]
\omega(z) ={}&
-1
+ \frac{
2(1+z)\gamma
\left(
\beta\mu
+ 3\gamma^{2}(1+\mu)
\left[1-(1+z)^{2}\gamma^{2}\right]
\right)
}
{
3
\left[1-(1+z)^{2}\gamma^{2}\right]^{3/2}
\left(
\beta
+ 3\gamma^{2}
\left[1-(1+z)^{2}\gamma^{2}\right]
\right)
}.
\end{aligned}
\end{equation}

\vspace{-0.6em}

\begin{equation}\label{eq:prhoomegam2}
\begin{aligned}
p(z) ={}&
-\frac{2^{-\mu} 3^{-1-\mu} n^{2(1+\mu)}}
{\alpha(1+2\mu)\,\eta^{2(1+\mu)}
\left[1+(1+z)^{2n}\right]^{1+\mu}}
\Bigg[
\frac{2\eta (1+z)^{2n-1}}{\sqrt{1+(1+z)^{2n}}}
\left(
\beta \mu
+ \frac{3\eta^2 (1+\mu)
\left[1+(1+z)^{2n}\right]}{n^2}
\right)
\\[-0.3em]
&\qquad
+ \frac{3\eta^2 \left[1+(1+z)^{2n}\right]}{n^2}
\left(
\beta
+ \frac{3\eta^2 \left[1+(1+z)^{2n}\right]}{n^2}
\right)
\Bigg],
\\[0.6em]
\rho(z) ={}&
\frac{
6^{-\mu} n^{2\mu} \eta^{-2\mu}
\left[1+(1+z)^{2n}\right]^{-\mu}
}
{\alpha(1+2\mu)}
\left(
\beta
+ \frac{3\eta^2}{n^2}
\left[1+(1+z)^{2n}\right]
\right),
\\[0.6em]
\omega(z) ={}&
-1
-\frac{n^2}
{3\eta^2
\left[1+(1+z)^{2n}\right]
\left(
\beta
+ \frac{3\eta^2 \left[1+(1+z)^{2n}\right]}{n^2}
\right)} \times
\frac{
2\eta (1+z)^{2n-1}
\left(
\beta \mu
+ \frac{3\eta^2 (1+\mu)
\left[1+(1+z)^{2n}\right]}{n^2}
\right)}
{\sqrt{1+(1+z)^{2n}}}.
\end{aligned}
\end{equation}
\section*{Acknowledgement}  
The author gratefully acknowledges that this research was conducted without internal or external funding support.
\end{widetext}
\bibliographystyle{utphys}
\bibliography{biblio}

\end{document}